\documentclass[12pt]{article}

\title{Geometry and Entanglement of Super-Qubit Quantum States}

\author{Oktay K Pashaev and Aygul Kocak\\
Department of Mathematics \\ Izmir Institute of Technology \\ Izmir 35430, Turkey}

\begin{document}

\maketitle              

\begin{abstract}
   We introduce the  super-qubit quantum state, determined by superposition of the zero and the one super-particle states, which can be represented by points on the super-Bloch sphere. 
In contrast to the one qubit case, the one super-particle state is  characterized by points in extended complex plain, equivalent to another super-Bloch sphere. Then, geometrically, 
the super-qubit quantum state is represented by two unit spheres, or the direct product of two Bloch spheres.  
By using the displacement operator, acting on the super-qubit state as the reference state, we construct the super-coherent states, becoming eigenstates of the super-annihilation operator,
and characterized by three complex numbers, the displacement parameter and stereographic projections of two super-Bloch spheres.
The states are fermion-boson entangled, and the concurrence of states is the product of two concurrences, corresponding to two Bloch spheres. We show geometrical meaning of concurrence 
as distance from point-state on the sphere to vertical axes - the radius of circle at horizontal plane through the point-state. Then, probabilities of collapse to the north pole state and to
the south pole state are equal to half-distances from vertical coordinate of the state to corresponding points at the poles. For complimentary fermion number operator, we get the
flipped  super-qubit state and corresponding super-coherent state, as eigenstate of 
transposed  super-annihilation operator. The infinite set of Fibonacci oscillating circles in complex plain, and corresponding set of quantum states with uncertainty relations as the ratio of two Fibonacci 
numbers, and in the limit at infinity becoming the Golden Ratio uncertainty, is derived.
\end{abstract}

Keywords:{qubit, super-qubit, supersymmetry, super-coherent states, super-Bloch sphere, entanglement}


 %
\section{Introduction}
 The unit of quantum information is called the qubit  and it is defined as superposition
\begin{equation}
|\theta, \phi\rangle = \cos \frac{\theta}{2} |0\rangle + \sin \frac{\theta}{2} e^{i\phi} |1\rangle
\end{equation}
of two basis states 
\begin{equation}
|0\rangle = \left( \begin{array}{c} 1 \\ 0    \end{array} \right), \hskip1cm |1\rangle = \left( \begin{array}{c} 0 \\ 1    \end{array} \right).
\end{equation}
 The states are eigenstates of the fermionic number operator
\begin{equation}
N_f = \left(  \begin{array}{cc} 0 & 0 \\ 0 & 1       \end{array}     \right),
\end{equation}
so that $N |0\rangle  = 0 |0\rangle $, $N |1\rangle  = 1 |1\rangle $. The qubit  can be realized by two level quantum system, such as spin $1/2$ or single fermion \cite{Benenti}.
An interaction of two level quantum system with bosons, then is described by  the fermion and the boson degrees of freedom. The natural question is, could we define
the unit of quantum information in the fermion-boson system. In present paper we introduce the analog of qubit in such system, which we call the supersymmetric qubit or shortly, the super-qubit. 
For simplicity, we will work with 
supersymmetric quantum oscillator \cite{Zypman} and with super-quanta or super-particles. By introducing the super-number operator ${\cal N}$, counting total number of superparticles 
in given state (without distinguishing fermions from bosons), we consider superposition of the zero $|0\rangle_{SS}$ and the one $|1\rangle_{SS}$ super-particle states, which natural to call 
as the super-qubit unit of quantum information. This superposition can be parametrized by two angles on the unit sphere, which we call the super-Bloch sphere. 
The specific difference with qubit case is that the one superparticle state $|1\rangle_{SS}$ is not unique. It is
in a superposition of one boson and one fermion states, determined by one complex number. By stereographic projection of the extended complex plane, 
the state $|1\rangle_{SS}$ can be parametrized also by points on the unit sphere. It means that in contrast with qubit state, the super-qubit state
requires extra parameters and it is characterized by two spheres. Due to this, information contents of super-qubit state becomes more reach and the states can be characterized by
entanglement of fermionic and bosonic states. Such entanglement was subject of study in several papers from different points of view \cite{R1}, \cite{R2}, \cite{R3}, \cite{R4}, \cite{P5}.

\section{Supersymmetric   Oscillator}

   The bosonic creation and annihilation operators $a^+$, $a$, (where the commutator is  $[a, a^\dagger] =I$), and the number operator $N = a^\dagger a$, $N+I = a a^\dagger $, are acting in 
the Fock space $H_b$, with $n$-particle states $\{|n\rangle\}$ as a basis. The states of supersymmetric quantum oscillator from $H_f \otimes H_b$ are characterized by
number of fermions $n_f = 0,1$ and number of bosons $n_b = 0,1,2,...$. The basis states are tensor products of fermion and boson states
$|0\rangle_f \otimes |n\rangle$ and $|1\rangle \otimes |n\rangle$. Then, an arbitrary state can be represented by superposition
\begin{equation}
|\Psi\rangle = \sum^\infty_{n=0} c_{0 n} |0\rangle_f \otimes |n\rangle + \sum^\infty_{n=0} c_{1 n} |1\rangle_f \otimes |n\rangle.\label{state}
\end{equation}
The super-number operator, counting total number of super-particles (fermions and bosons) in a state, is defined as
\begin{equation}
{\cal N} = \left(  \begin{array}{cc} N & 0 \\ 0 & N+1       \end{array}     \right) = N_f \otimes I_b + I_f \otimes N.
\end{equation}
It gives ${\cal N} |0\rangle_f \otimes |n\rangle = (0+n)|0\rangle_f \otimes |n\rangle $, ${\cal N} |1\rangle_f \otimes |n\rangle = (1+n)|1\rangle_f \otimes |n\rangle $.
Then, the states $|0\rangle_f \otimes |n\rangle$ and $|1\rangle_f \otimes |n-1\rangle$ contain the same number $n$ of superparticles. Moreover, any 
superposition of these states also have $n$ superparticles. Hence, the  normalized generic $n$ super-number state 
\begin{equation}
|n, \zeta\rangle = \frac{|0\rangle_f \otimes |n\rangle_b + \zeta |1\rangle_f \otimes |n-1\rangle_b}{\sqrt{1 + |\zeta|^2}} = \frac{1}{\sqrt{1 + |\zeta|^2}}
\left( \begin{array}{c} |n\rangle \\ \zeta |n-1\rangle   \end{array} \right),\label{nstate}
\end{equation} 
where $\zeta$ is an arbitrary complex number,
is the eigenstate of super-number operator ${\cal N} |n, \zeta\rangle  = n |n, \zeta\rangle $.
The origin $\zeta =0$ of complex plane $\zeta$  corresponds to $n$ pure bosons, while the infinity $\zeta = \infty$ in the extended complex plane, to the one fermion and $n-1$ bosons.
By stereographic projection, the extended complex plane can be projected to the unit sphere by formula
\begin{equation}
\zeta = \tan \frac{\theta_1}{2} e^{i\phi_1},\label{stereo1}
\end{equation}
so that the state becomes
\begin{equation}
|n, \theta_1, \phi_1\rangle =
\cos \frac{\theta_1}{2}\left( \begin{array}{c} |n\rangle \\  0 \end{array} \right) + \sin \frac{\theta_1}{2} e^{i\phi_1}\left( \begin{array}{c} 0 \\  |n-1\rangle \end{array} \right),
\label{stereo2}
\end{equation} 
where $0 \le \theta_1 \le \pi$, $0 \le \phi_1 \le 2\pi$ are angles on the sphere.

\subsection{Entanglement of Super-Number State}
 The super-number  states are fermion-boson entangled. If the state (\ref{state}) is the product state $|\phi\rangle_f \otimes |\xi\rangle_b$, then it is fermion-boson separable.
If not, then the state is entangled. By splitting the state (\ref{state}) in the form
\begin{equation}
|\Psi\rangle = |0\rangle_f \otimes |\psi_0\rangle + |1\rangle_f \otimes |\psi_1\rangle \label{state1}
\end{equation}
we can see that it is determined by two states in Fock space of bosons
\begin{equation}
|\psi_0\rangle = \sum^\infty_{n=0} c_{0 n}  |n\rangle, \hskip1cm |\psi_1\rangle = \sum^\infty_{n=0} c_{1 n}  |n\rangle.
\end{equation}
If these states are linearly dependent, then the state (\ref{state}) is separable. To characterize the fermion-boson entanglement of the state we 
use the reduced density matrix approach as in \cite{P5}. By taking trace of density matrix $\rho = |\Psi \rangle \langle \Psi|$ according to fermion degrees of 
freedom, we get the reduced bosonic density matrix
\begin{equation}
\rho_b = Tr_f \rho = |\psi_0\rangle \langle \psi_0| + |\psi_1\rangle \langle \psi_1|.
\end{equation}
The linear entropy is counting deviation of $Tr \rho^2_b$ from 1 and it is called as the concurrence square. The calculation gives \cite{P5}
\begin{equation}
\frac{1}{2} C^2 = 1 - Tr \rho^2_b = 2 \left|  \left(  \begin{array}{cc} \langle \psi_0|\psi_0 \rangle & \langle \psi_0|\psi_1 \rangle \\ \langle \psi_1|\psi_0 \rangle &  \langle \psi_1|\psi_1 \rangle     \end{array}     \right)                 \right|,
\end{equation}
and as follows
the square of concurrence for state (\ref{state1}) is given by the Gram determinant for inner products in Fock space
\begin{equation}
C^2 =  4 \left|  \left(  \begin{array}{cc} \langle \psi_0|\psi_0 \rangle & \langle \psi_0|\psi_1 \rangle \\ \langle \psi_1|\psi_0 \rangle &  \langle \psi_1|\psi_1 \rangle     \end{array}     \right)                 \right|. \label{Gram}
\end{equation}
For $n$-super-particle state (\ref{nstate}) the concurrence is not dependent on $n$ and it is equal
	\begin{equation}
	C = \frac{2 |\zeta|}{1 + |\zeta|^2}.\label{concurrence}
	\end{equation}
	This implies that for $\zeta = 0$ and $\zeta = \infty$, the states are separable ($C =0$), while on the unit circle $|\zeta|^2 =1$, the states are maximally
	entangled ($C=1$).
	\subsubsection{One Super-Particle State}
	For $n=1$ we have the state
	\begin{equation}
|1, \zeta\rangle = \frac{|0\rangle_f \otimes |1\rangle_b + \zeta |1\rangle_f \otimes |0\rangle_b}{\sqrt{1 + |\zeta|^2}} = \frac{1}{\sqrt{1 + |\zeta|^2}}
\left( \begin{array}{c} |1\rangle \\ \zeta |0\rangle   \end{array} \right),\label{1state}
\end{equation} 
	which is fermion-boson entangled one super-particle state. The level of entanglement is determined by formula (\ref{concurrence}). In terms of the stereographic projection 
	(\ref{stereo1}), (\ref{stereo2}), it gives the concurrence 
	\begin{equation}
	C = \sin \theta_1.
	\end{equation}
	This formula provides simple geometrical meaning of concurrence on the unit sphere.  Indeed, the concurrence corresponding to state $|1, \theta_1, \phi_1\rangle$ is equal
	to distance from point $(\theta_1,\phi_1)$ on the sphere,  from the vertical axis. 
	It is equivalent to the radius of circle in horizontal plane, intersecting vertical axis at $z = \sqrt{1-C^2} = \cos \theta_1$.
	The last formula gives the von Neumann entropy as function of $z$ only
	\begin{equation}
	E = -\frac{1}{2} \log_2 \frac{1-z^2}{4} - \frac{z}{2} \log_2 \frac{1+z}{1-z}.
	\end{equation}
	The probabilities of collapse to states  $|0\rangle_f \otimes |1\rangle$ and $|1\rangle_f \otimes |0\rangle$,
	corresponding to poles on the sphere are equal correspondingly
	\begin{equation}
	p_0 = \cos^2 \frac{\theta}{2} = \frac{1+z}{2}, \hskip1cm p_1 = \sin^2 \frac{\theta}{2} = \frac{1-z}{2}.
	\end{equation}
	This means that geometrically, the probabilities are equal to half-distances from vertical projection of the state to the north and the south poles.

	For the one super-number  state (\ref{1state}), ${\cal N} |1, \zeta\rangle = 1 |1, \zeta\rangle$. In addition, for $n=0$ we have the 
	separable state
	\begin{equation}
|0, \zeta\rangle = |0\rangle_f \otimes |0\rangle_b   =  
\left( \begin{array}{c} |0\rangle \\ 0  \end{array} \right),\label{0state}
\end{equation} 
	satisfying ${\cal N} |0, \zeta\rangle = 0 |0, \zeta\rangle$ and orthogonal to the first one $\langle 0, \zeta| 1,\zeta \rangle =0$.
	The states are related by quantum gate, in the form of the creation supersymmetric operator
	\begin{equation}
	|1, \zeta\rangle = \frac{1}{\sqrt{1 + |\zeta|^2}}
\left( \begin{array}{c} |1\rangle \\ \zeta |0\rangle   \end{array} \right) = \frac{1}{\sqrt{1+ |\zeta|^2}} \left(  \begin{array}{ccc} a^\dagger & &0 \\  & & \\
	\zeta & &a^\dagger       \end{array}     \right) \left( \begin{array}{c} |0\rangle \\ 0  \end{array} \right).
	\end{equation}
	As we will see later, the Hermitian conjugate of this operator in the form of the supersymmetric annihilation operator (\ref{refcondition}), determines the supercoherent states as the eigenstates.  

\subsubsection{ Super-Qubit State}
	By taking superposition of $n=0$ and $n=1$ states we get the super-qubit quantum state
	\begin{equation}
	|\theta, \phi, \zeta\rangle = \cos \frac{\theta}{2} |0, \zeta\rangle + \sin \frac{\theta}{2} e^{i\phi} |1, \zeta\rangle,\label{superqubit}
	\end{equation}
	or in explicit form 
	\begin{equation}
	|\theta, \phi, \zeta\rangle =
	\cos \frac{\theta}{2} \left( \begin{array}{c} |0\rangle \\ 0  \end{array} \right) +
	\sin \frac{\theta}{2} e^{i\phi} \frac{1}{\sqrt{1 + |\zeta|^2}}
\left( \begin{array}{c} |1\rangle \\ \zeta |0\rangle   \end{array} \right), \label{superqubit1}
	\end{equation}
	characterized by two real $\theta$, $\phi$ and one complex $\zeta$ parameters.
	For this state, the first two parameters are angles on the unit sphere, which we call the super-Bloch sphere. Indeed, the north pole of the sphere
	at $\theta = 0$ corresponds to the zero number of super-particles in state $|0,0,\zeta\rangle$, while the south pole at $\theta =\pi$, to the one super-particle
	in state $|\pi,0,\zeta \rangle$. 
	Then, any point on the sphere gives superposition of these two states, with different level of entanglement.
	Probabilities to measure the basis states are equal
	\begin{eqnarray}
	p_0=|\langle 0, \zeta| \theta, \phi, \zeta\rangle|^2 = \cos^2 \frac{\theta}{2}, \hskip0.5cm p_1=|\langle 1, \zeta| \theta, \phi, \zeta\rangle|^2 = \sin^2 \frac{\theta}{2}.
	\end{eqnarray}
	The super-qubit state  (\ref{superqubit1}) is natural generalization of the fermionic or bosonic one qubit states and
in the limiting cases it reduces to separable qubit states:
	1. For $\zeta = 0$, the state is separable one qubit bosonic state
\begin{eqnarray}
	|\theta, \phi, 0\rangle =
	\cos \frac{\theta}{2} \left( \begin{array}{c} |0\rangle \\ 0  \end{array} \right) +
	\sin \frac{\theta}{2} e^{i\phi} 
\left( \begin{array}{c} |1\rangle \\ 0  \end{array} \right) = |0\rangle_f \otimes |\theta, \phi\rangle_b, 
	\end{eqnarray}
	2. For $\zeta = \infty$ it is separable one qubit fermionic state
	\begin{eqnarray}
	|\theta, \phi, \infty\rangle =
	\cos \frac{\theta}{2} \left( \begin{array}{c} |0\rangle \\ 0  \end{array} \right) +
	\sin \frac{\theta}{2} e^{i\phi} 
\left( \begin{array}{c} 0 \\ |0\rangle  \end{array} \right) = |\theta, \phi\rangle_f \otimes |0\rangle_b.
	\end{eqnarray}

	In general, the state is entangled. 
		For $\zeta = e^{i\gamma}$, so that $|\zeta|^2 =1$, the super-qubit state  (\ref{superqubit1}) reduces to the one 
	\begin{equation}
	|\theta, \phi, e^{i\gamma}\rangle =
	\cos \frac{\theta}{2} \left( \begin{array}{c} |0\rangle \\ 0  \end{array} \right) +
	\sin \frac{\theta}{2} e^{i\phi} \frac{1}{\sqrt{2}}
\left( \begin{array}{c} |1\rangle \\ e^{i\gamma}|0\rangle   \end{array} \right), 
	\end{equation}
	in particular cases, for $\gamma = 0$ and $\pi$, giving the pair of states 
	\begin{equation}
	|\theta, \phi, \pm\rangle =
	\cos \frac{\theta}{2} \left( \begin{array}{c} |0\rangle \\ 0  \end{array} \right) +
	\sin \frac{\theta}{2} e^{i\phi} \frac{1}{\sqrt{2}}
\left( \begin{array}{c} |1\rangle \\ \pm|0\rangle   \end{array} \right), 
	\end{equation}
	considered as the reference states in \cite{P5}. Then, the corresponding one super-particle states
	\begin{equation}
	|\pi, \phi, e^{i\gamma}\rangle = |1, e^{i\phi}\rangle = 
	\frac{1}{\sqrt{2}}
\left( \begin{array}{c} |1\rangle \\ e^{i\gamma}|0\rangle   \end{array} \right), 
	\end{equation}
	are maximally entangled states with $C =1$.

\subsubsection{Entanglement of Super-Qubit State}

	The super-qubit state (\ref{superqubit}) is fermion-boson entangled state with concurrence given by the product formula
	of two concurrences
	\begin{equation}
	C = \sin^2 \frac{\theta}{2} \, \frac{2 |\zeta|}{1 + |\zeta|^2}.
	\end{equation}
	For $\zeta =0$ and $\zeta =1$, the super-qubit state is separable and $C =0$. For $|\zeta|=1$ the concurrence is equal
	\begin{equation}
	C = \sin^2 \frac{\theta}{2}, 
	\end{equation}
	as in the special case of super-qubit state considered in \cite{P5}.
	\section{Displacement Operator and Super-Qubit States}
	\subsection{Super-Qubit Annihilation operator}
		The super-qubit state (\ref{superqubit}) is the reference state, annihilated by super-annihilation operator
	\begin{equation}
	A_{-1/\zeta} = \left(  \begin{array}{ccc} a & &-\frac{1}{\zeta} \\  & & \\
	0 & &a       \end{array}     \right),\hskip1cm A_{-1/\zeta} |\theta, \phi, \zeta\rangle =0.\label{refcondition}
	\end{equation}
	This follows from observation that for basis states $A_{-1/\zeta} |0, \zeta\rangle =0$ and $A_{-1/\zeta} |1, \zeta\rangle =0$. 
	The operators $A_{-1/\zeta}$ and  $A^\dagger_{-1/\zeta}$ 
	satisfy following commutation relations with super-number operator
	\begin{equation}
	[{\cal N}, A_{-1/\zeta}] = -A_{-1/\zeta}, \hskip0.5cm [{\cal N}, A^\dagger_{-1/\zeta}] = + A^\dagger_{-1/\zeta},
	\end{equation}
	\begin{equation}
	[A_{-1/\zeta}, A^\dagger_{-1/\zeta}] = I + \frac{1}{|\zeta|^2} \sigma_3 \otimes I_b.
	\end{equation}
	In the limit $\zeta = \infty$ the operator  $A_{-1/\zeta}$ is just the direct product $I_f \otimes a$ and the algebra becomes identical to the one for pure bosonic
	operators.
	
	\subsection{Displacement Operator}
	In a similar way as in \cite{P5}, we introduce the displacement operator acting in $H_f \otimes H_b$ space,
	\begin{equation}
	{\cal D}(\alpha) = I_f \otimes D(\alpha) = I_f \otimes e^{\alpha a^\dagger - \bar\alpha a}
	\end{equation}
	with following commutator
		\begin{equation}
	[A_{-1/\zeta}, {\cal D}(\alpha)] = A_{-1/\zeta} {\cal D}(\alpha) - {\cal D}(\alpha) A_{-1/\zeta} = \alpha {\cal D}(\alpha).\label{commutator}
\end{equation}
Then, the supersymmetric coherent state, associated with the super-qubit reference state (\ref{superqubit}) is defined as
\begin{equation}
|\alpha, \theta, \phi, \zeta \rangle = {\cal D}(\alpha) |\theta, \phi, \zeta \rangle, \label{coherent}
\end{equation}
and it can be represented in the form
\begin{equation}
|\alpha, \theta, \phi, \zeta \rangle = 
	\cos \frac{\theta}{2} \left( \begin{array}{c} |0, \alpha\rangle \\ 0  \end{array} \right) +
	\sin \frac{\theta}{2} e^{i\phi} \frac{1}{\sqrt{1 + |\zeta|^2}}
\left( \begin{array}{c} |1, \alpha\rangle \\ \zeta |0, \alpha\rangle   \end{array} \right), \label{superqubitcoherent}
	\end{equation}
	determined by displaced $|0\rangle$ and $|1\rangle$ states, $|0, \alpha\rangle = D(\alpha) |0\rangle$, $|1, \alpha\rangle = D(\alpha) |1\rangle$.
	As easy to show, 
	the super-coherent states (\ref{superqubitcoherent}) are eigenstates of super-annihilation operator $A_{-1/\zeta}$,
\begin{equation}
A_{-1/\zeta} |\alpha, \theta, \phi, \zeta \rangle = \alpha |\alpha, \theta, \phi, \zeta \rangle.\label{eigenvalue}
\end{equation}
The proof is following.
Applying the operator (\ref{refcondition}) to the state  and using commutation relation (\ref{commutator})  we get
\begin{equation}
A_{-1/\zeta} |\alpha, \theta, \phi, \zeta \rangle = A_{-1/\zeta} {\cal D}(\alpha) |\theta, \phi, \zeta \rangle = {\cal D}(\alpha) A_{-1/\zeta} |\theta, \phi, \zeta \rangle
+ \alpha {\cal D}(\alpha) |\theta, \phi, \zeta \rangle.  
\end{equation}
The first term in r.h.s.,  due to (\ref{refcondition}),  vanishes and we obtain the desired result.

The above results show that similarly to Glauber coherent states, two definitions of super-coherent states (\ref{coherent}) and (\ref{eigenvalue})  are equivalent.
In particular case $\zeta =\pm 1$, the state (\ref{superqubitcoherent}) reduces to the one, introduced in \cite{P5}.
\subsubsection{Concurrence of Super-Coherent States}

	To calculate entanglement of state (\ref{superqubitcoherent}) we notice that
\begin{equation}
|\alpha, \theta, \phi, \zeta \rangle = {\cal D}(\alpha) |\theta, \phi, \zeta \rangle = |0\rangle_f \otimes D(\alpha) |\psi_0\rangle + |1\rangle_f \otimes D(\alpha) |\psi_1\rangle,
\end{equation}
where the reference state is $|\theta, \phi, \zeta \rangle = |0\rangle_f \otimes |\psi_0\rangle + |1\rangle_f \otimes |\psi_1\rangle$, and 
the super-coherent state
\begin{equation}
|\alpha, \theta, \phi, \zeta \rangle = |0\rangle_f \otimes |\psi_0, \alpha\rangle + |1\rangle_f \otimes |\psi_1, \alpha\rangle,
\end{equation}
is determined by two displaced Fock states $|\psi_0, \alpha\rangle$ and $|\psi_1, \alpha\rangle$.
Since the Fock states are connected by unitary transformation 
\begin{equation}
|\psi_0, \alpha \rangle = D(\alpha) |\psi_0\rangle, \hskip1cm |\psi_1, \alpha \rangle = D(\alpha) |\psi_1\rangle,
\end{equation}
the inner products will not depend on $\alpha$,
\begin{equation}
\langle \psi_i, \alpha | \psi_j, \alpha \rangle = \langle\psi_i | D^\dagger(\alpha) D(\alpha) |\psi_j\rangle = \langle\psi_i |\psi_j\rangle,
\end{equation} 
as well as the Gram determinant (\ref{Gram}) of the inner products and corresponding concurrence $C$. Thus,  we obtain the following result:
the concurrence (\ref{Gram}) for displaced state (\ref{state}), $|\Psi, \alpha \rangle = {\cal D}(\alpha) |\Psi\rangle $ is independent of $\alpha$ and it is equal to
the concurrence for the reference state $|\Psi\rangle$.
As a result, for the super-coherent state (\ref{superqubitcoherent}),
the concurrence coincides with the one for the super-qubit state (\ref{superqubit}) and it is equal
\begin{equation}
	C = \sin^2 \frac{\theta}{2} \, \frac{2 |\zeta|}{1 + |\zeta|^2}.
	\end{equation}
\section{Flipped Super-Qubits and Super-Coherent States}
The flipping operator we define as
\begin{equation}
{\cal X} = X \otimes I_b = \sigma_1 \otimes I_b.
\end{equation}
It acts on the fermion number operator in following way
\begin{equation}
{\cal X}     (  N_f \otimes I_b   )     {\cal X} = \bar N_f \otimes  I_b,
\end{equation}
where $\bar N_f = diag(1,0)$ corresponds to interchanging the number of fermions in states. 
Applying operator ${\cal X}$ to $n$- super-number state (\ref{nstate}) we get the flipped state
\begin{equation}
{\cal X}|n, \zeta\rangle = \frac{|1\rangle_f \otimes |n\rangle_b + \zeta |0\rangle_f \otimes |n-1\rangle_b}{\sqrt{1 + |\zeta|^2}} = \frac{1}{\sqrt{1 + |\zeta|^2}}
\left( \begin{array}{c} \zeta|n-1\rangle \\ |n\rangle   \end{array} \right).\label{fnstate}
\end{equation} 
	The flipped one super-particle state is equal
\begin{equation}
{\cal X}|1, \zeta\rangle = \frac{1}{\sqrt{1 + |\zeta|^2}}
\left( \begin{array}{c} \zeta|0\rangle \\ |1\rangle   \end{array} \right) = \frac{\zeta |0\rangle_f \otimes |0\rangle_b + |1\rangle_f \otimes |1\rangle_b}{\sqrt{1+|\zeta|^2}},
\label{f1state}
\end{equation}
and corresponding flipped super-qubit state is
\begin{equation}
	{\cal X}|\theta, \phi, \zeta\rangle =
	\cos \frac{\theta}{2} \left( \begin{array}{c} 0 \\ |0\rangle  \end{array} \right) +
	\sin \frac{\theta}{2} e^{i\phi} \frac{1}{\sqrt{1 + |\zeta|^2}}
\left( \begin{array}{c} \zeta|0\rangle \\  |1\rangle   \end{array} \right). \label{fsuperqubit1}
	\end{equation}
	It turns out that the concurrence for state $|\Psi\rangle$ and the flipped one ${\cal X} |\Psi\rangle$ is the same. 
	
	The proof is following. If the state $|\Psi\rangle$ is represented as
\begin{equation}
|\Psi\rangle = |0\rangle_f \otimes |\psi_0\rangle + |1\rangle_f \otimes |\psi_1\rangle,
\end{equation}
then the flipped state is
\begin{equation}
{\cal X}|\Psi\rangle = {\cal X}|0\rangle_f \otimes |\psi_0\rangle + {\cal X}|1\rangle_f \otimes |\psi_1\rangle = |1\rangle_f \otimes |\psi_0\rangle + |0\rangle_f \otimes |\psi_1\rangle.
\end{equation}
This implies that results of flipping for the state is in interchanging the Fock states $|\psi_0\rangle$ and $|\psi_1\rangle$ and the corresponding inner products. 
But, due to representation of concurrence by the Gram determinant (\ref{Gram}), it is clear that the determinant as well as the concurrence are invariant under such replacement.

Due to this result, 	the concurrence for flipped state (\ref{f1state}) is equal to the one for the state $|1,\zeta\rangle$,
\begin{equation}
C = \frac{2 |\zeta|}{1 + |\zeta|^2},
\end{equation}
and the concurrence for flipped super-qubit state (\ref{fsuperqubit1}) is equal to the one for the super-qubit state
\begin{equation}
C = \sin^2 \frac{\theta}{2}\,\frac{2 |\zeta|}{1 + |\zeta|^2}.\label{fconcurrence}
\end{equation}

Application of flipping gate to super-annihilation operator gives the transposed operator
\begin{equation}
{\cal X} A_{-1/\zeta} {\cal X} = A^T_{-1/\zeta} = \left(  \begin{array}{ccc} a & &0 \\  & & \\
	-\frac{1}{\zeta} & &a       \end{array}     \right),
\end{equation}
and the flipped super-qubit state is annihilated by this operator
\begin{equation}
A^T_{-1/\zeta} {\cal X}|\theta, \phi, \zeta\rangle =0.
\end{equation}
Then, the flipped super-coherent state is defined by action of displacement operator on flipped super-qubit state
 \begin{equation}
|\alpha, \theta, \phi, \zeta \rangle_{\cal X} = {\cal D}(\alpha) {\cal X}|\theta, \phi, \zeta \rangle. \label{fcoherent}
\end{equation}
	Since operators ${\cal D}(\alpha)$ and ${\cal X}$ commute, $[{\cal D}(\alpha), {\cal X}] =0$, we have the following result.

The flipped super-coherent state can be represented as
\begin{equation}
{\cal X}|\alpha, \theta, \phi, \zeta \rangle = 
	\cos \frac{\theta}{2} \left( \begin{array}{c} 0 \\ |0, \alpha\rangle  \end{array} \right) +
	\sin \frac{\theta}{2} e^{i\phi} \frac{1}{\sqrt{1 + |\zeta|^2}}
\left( \begin{array}{c} \zeta |0, \alpha\rangle \\ |1, \alpha\rangle   \end{array} \right). \label{fsuperqubitcoherent}
	\end{equation}
	It is the eigenstate of $A^T_{-1/\zeta}$ operator,
	\begin{equation}
A^T_{-1/\zeta} {\cal X}|\alpha,\theta, \phi, \zeta\rangle =\alpha {\cal X}|\alpha,\theta, \phi, \zeta\rangle,
\end{equation}
having the same concurrence (\ref{fconcurrence}) as for the super-qubit state.

\section{Uncertainty Relations and Fibonacci Sequence} 
The coordinate and momentum operators, defined by
\begin{equation}
X = I_f \otimes \frac{a + a^\dagger}{\sqrt{2}}, \hskip1cm P = I_f \otimes i\frac{a^\dagger - a}{\sqrt{2}},
\end{equation}
transform by the dislpacement operator as
\begin{equation}
{\cal D}^\dagger(\alpha) X {\cal D}(\alpha) = X + \sqrt{2} \Re \alpha, \hskip0.5cm {\cal D}^\dagger(\alpha) P {\cal D}(\alpha) = P + \sqrt{2} \Im \alpha.
\end{equation}
Then, the average values of $X$ and $P$ operators in super-coherent states are 
\begin{eqnarray}
\langle \alpha, \theta, \phi, \zeta |X | \alpha, \theta, \phi, \zeta \rangle &=& \sqrt{2} \Re \alpha + \frac{\sin\theta \cos \phi}{2 \sqrt{1 + |\zeta|^2}}, \\
\langle \alpha, \theta, \phi, \zeta |P | \alpha, \theta, \phi, \zeta \rangle &=& \sqrt{2} \Im \alpha + \frac{\sin\theta \sin \phi}{2 \sqrt{1 + |\zeta|^2}}.
\end{eqnarray}

The classical values of coordinate and momentum we denote as  $x_c = \sqrt{2} \Re\alpha$, $y_c = \sqrt{2} \Im \alpha$ in complex plane $\alpha$.
The spherical coordinates on the super-Bloch sphere are 
\begin{equation}
x = \sin\theta \cos \phi, \,\,\, y = \sin \theta \sin \phi, \,\,\, z = \cos \theta,
\end{equation}
so that $2 \sin^2 \frac{\theta}{2} = 1-z$, and $x^2 + y^2 + z^2 =1$.
Then, the  average values of coordinate and momentum are represented by following projections
\begin{eqnarray}
\langle \alpha, \theta, \phi, \zeta |X | \alpha, \theta, \phi, \zeta \rangle &=& x_c + \frac{x}{2 \sqrt{1 + |\zeta|^2}}, \\
\langle \alpha, \theta, \phi, \zeta |P | \alpha, \theta, \phi, \zeta \rangle &=& y_c + \frac{y}{2 \sqrt{1 + |\zeta|^2}}.
\end{eqnarray}

The dispersions or the variance of $X$ and $P$ operators are defined as $(\Delta X)^2 = \langle X^2 \rangle - \langle X \rangle^2  $, $(\Delta P)^2 = \langle P^2 \rangle - \langle P \rangle^2  $.
Then, by direct calculation we can show that
dispersions of $X$ and $P$ operators in super-coherent states are not dependent on $\alpha$ and are equal
\begin{eqnarray}
(\Delta X)^2 &=& \frac{1}{2}\left(1 + \frac{2\sin^2 \frac{\theta}{2} - \sin^2 \theta \cos^2 \phi}{ {1 + |\zeta|^2}}\right), \\
(\Delta P)^2 &=& \frac{1}{2}\left(1 + \frac{2\sin^2 \frac{\theta}{2} - \sin^2 \theta \sin^2 \phi}{ {1 + |\zeta|^2}}\right),
\end{eqnarray}
 or in terms of the Cartesian coordinates of the super-Bloch sphere 
\begin{eqnarray}
(\Delta X)^2 &=& \frac{1}{2}\left(1 + \frac{1 - z - x^2}{ {1 + |\zeta|^2}}\right), \\
(\Delta P)^2 &=& \frac{1}{2}\left(1 + \frac{1 -z - y^2}{ {1 + |\zeta|^2}}\right).
\end{eqnarray}

For angle $\phi = \frac{\pi}{4}$, the coordinates $x=y$ and dispersions are equal. So that for the corresponding state in equatorial plane $\theta = \frac{\pi}{2}$, 
\begin{equation}
(\Delta X)^2 = (\Delta P)^2 = \frac{1}{2}\left( 1 + \frac{1}{2 (1+ |\zeta|^2)}        \right).
\end{equation}
From this formula for states with $|\zeta| = 1$ we have dispersions
\begin{equation}
(\Delta X)^2 = (\Delta P)^2 = \frac{5}{8} = \frac{F_5}{F_6},
\end{equation}
as ratio of two Fibonacci numbers.
This suggests to construct the infinite sequence of quantum states with dispersion as the ratio of Fibonacci numbers for any $n$.
Indeed, the sequence of circles 
\begin{equation}
|\zeta_n|^2 = \frac{F_{n-1}}{F_{n-2}} - \frac{1}{2}
\end{equation}
in complex plane $\zeta$, determines dispersions
\begin{equation}
(\Delta X_n)^2 = (\Delta P_n)^2 = \frac{F_n}{F_{n+1}},
\end{equation}
and uncertainty relations
\begin{equation}
\Delta X_n \Delta P_n = \frac{F_n}{F_{n+1}},
\end{equation}
by the Golden sequence $\varphi_n = F_{n+1}/F_n$.
The radius square of circles $|\zeta_n|^2$ is Fibonacci oscillating around the value, corresponding to the limit  $n \rightarrow \infty$,
\begin{equation}
|\zeta_\infty|^2 = \varphi - \frac{1}{2},
\end{equation}
where $\varphi$ is the Golden Ratio.
In this limit we have the Golden dispersions 
\begin{equation}
(\Delta X_\infty)^2 = (\Delta P_\infty)^2 = \frac{1}{\varphi},
\end{equation}
and the Golden uncertainty relation
\begin{equation}
\Delta X_\infty \Delta P_\infty = \frac{1}{\varphi}.
\end{equation}

\section{Conclusions}
 We have introduced the super-qubit quantum state as an extension of usual qubit state to supersymmetric case. The state is superposition of 
the zero and the one super-particle states, determined by two real parameters on the unit sphere, but in contrast with the usual qubit state, the one 
super-particle state  is also determined by its own two real parameters on a unit sphere or in the corresponding extended complex plane. The super-qubit state
is annihilated by related super-annihilation operator, and the displaced super-qubit state in the form of the super-coherent state, becomes eigenstate of this operator.
The fermion-boson entanglement in super-coherent state is the same as in super-qubit state and it is determined by two complex parameters. 
By flipping transformation of super-qubit state, the flipped  super-coherent states were derived and it was shown that they have the same level of
entanglement. We also have introduced the infinite set of Fibonacci oscillating circles in complex plain, which describe quantum states with uncertainty relations as the ratio of two Fibonacci 
numbers. In the limit $n \rightarrow \infty$, it becomes the Golden Ratio uncertainty.
The super-qubit states and the super-coherent states introduced in the present paper are generalizations of the states derived in \cite{P5}. In the limit, 
when the one super-particle state is maximally entangled state, our states reduce to the ones in \cite{P5}. 
By using higher super-particle states it is possible to construct super-qudit states as an extension of the super-qubit and this idea  is under investigation
now. 

This work was supporting by BAP project 2022IYTE-1-0002.

PS. When this paper was completed, we have been kindly informed by S. Duplij that another type of super-qubit states in terms of the superspace with Grassman variables 
was introduced in \cite{B} and explored in \cite{D}. It would be interesting to apply the displacement operator to create the supersymmetric coherent states in this approach 
as well.

%

\end{document}